\documentclass[11pt,dvips]{article}
\usepackage{epsfig}
\usepackage{wrapfig}
\begin{document}
\begin{center}

{\LARGE \bf The Dependence of the Age Parameter from EAS Size and 
Zenith Angle of Incidence}
\end{center}

\begin{center}
{\bf A.A. Chilingarian, 
G.V. Gharagyozyan\footnote{corresponding author: 
gagik@crdlx5.yerphi.am}, S.S. Ghazaryan, G.G. Hovsepyan, \\
E.A. Mamidjanyan, L.G. Melkumyan, S.H. Sokhoyan} \\
{\it Yerevan Physics Institute, Cosmic Ray Division, Armenia}
\end{center}

\vspace{1ex}
\begin{center}
\begin{minipage}[c]{5.5in}
The quality of the MAKET-ANI detector installation in view of the 
uniformity of the registration efficiency is demonstrated. 
Based on a data sample collected by the MAKET-ANI array in the 
period of June 1997 - March 1999, the dependencies of the age 
parameter on the zenith angle and the EAS size $(10^5-10^7)$ are 
studied. The variation of the age parameter with the shower size 
can be approximately related to the elongation rate.
\end{minipage}
\end{center}
\vspace{1ex}

\section{Introduction}
The lateral distribution of the charged particle component of 
extensive air showers (EAS) carries information about the height 
of maximum of the EAS development. In NKG type parameterizations 
of the lateral distribution this information is associated 
with the so-called age parameter s, originally introduced by 
the analytic description of purely electromagnetic cascades 
for characterizing the actual stage of the EAS development. \\
In EAS experiments this parameter is usually extracted from 
fitting the distribution measured on observed level, assuming 
that this lateral parameter reflects the actual longitudinal 
EAS stage. Investigations of the parameter s have been performed 
on various altitudes, with the aim to gain information on the 
longitudinal EAS development and on the composition of primary 
cosmic rays [1-7]. For example, from the 
analysis of the zenith angle dependence of the average value of  
s it has been concluded that the mass composition gets either 
heavier primary energies larger than $10^{15}\,$eV or the 
multiplicity of secondary particle production in hadronic 
interactions is unexpectedly increasing. \\
In the present contribution experimental age distributions, 
dependent on the zenith angle $\theta$ of EAS incidence and of 
the shower size $N_e$ as extracted from an actual data set  
of the MAKET-ANI array, are communicated. As compared to earlier  
results \cite{AH-avak} the statistical accuracy of the data is 
considerably improved thanks to various modernizations of 
the installation \cite{AH-gagik}. The variation of the age 
parameter with the observation depth X is considered by a 
simplified approach.

\section{Some characteristics of the data selection}
With an effective running time of ca. 8000 h the array triggered 
for $2.6\cdot10^6$ showers. From this set 177066 showers have 
been selected with following criteria: $N_e\geq1\times10^5$,
$\theta<45^o$, $0.3\leq s \leq 1.7$. The procedures of data 
selection and further analyses are given in Ref. \cite{AH-gagik}.
The effective area for EAS registration, varying from 
$28\cdot14$m$^2$ for $N_e\geq10^5$ to $64\cdot32m^2$ for 
$N_e\geq10^6$. With Monte Carlo simulations and experimental 
considerations of the angular accuracy following uncertainties 
of the reconstructed EAS parameters were obtained: core location: 
$\delta R \simeq1.5\,$m,  $\delta N_e \simeq 15\%$ for $N_e<10^6$,
$\delta N_e \simeq 10\%$ for $N_e>10^6$, $\delta s \simeq 7\%$,
$\delta\theta<1.5^o$ and $\delta\varphi<5^o$. \\
Figures 1 and 2 display the good uniformity of the EAS registration; 
the maximum intensity results %
\begin{figure}
\begin{minipage}[b]{0.48\linewidth}
\epsfig{file=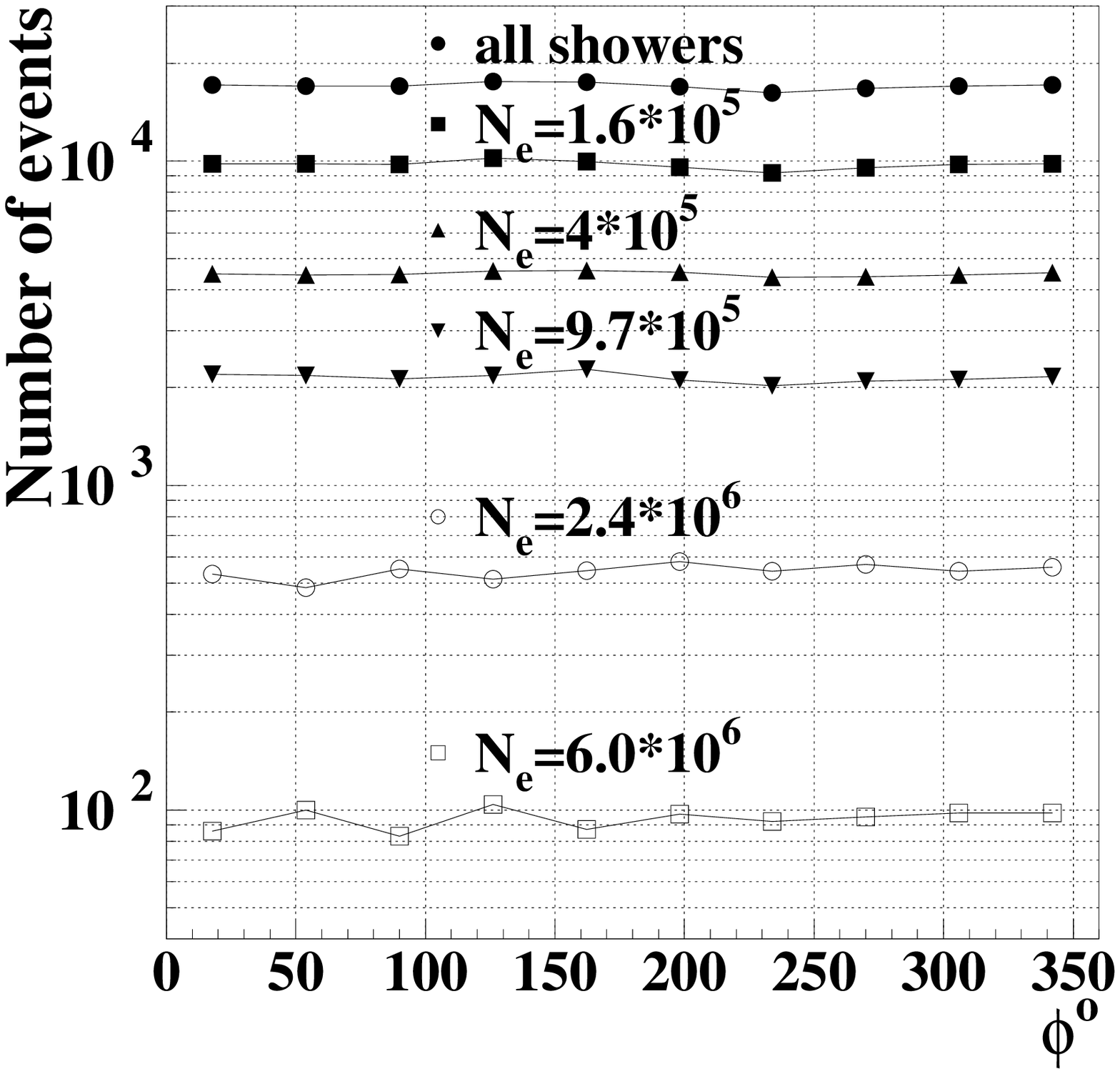,width=8.cm,height=8.cm}
\vspace{-1.5cm}
\caption{\it{Distributions of the EAS azimuth angles 
$\varphi$ for different EAS sizes.}}
\end{minipage}
\hspace*{0.4cm}
\begin{minipage}[b]{0.48\linewidth}
\epsfig{file=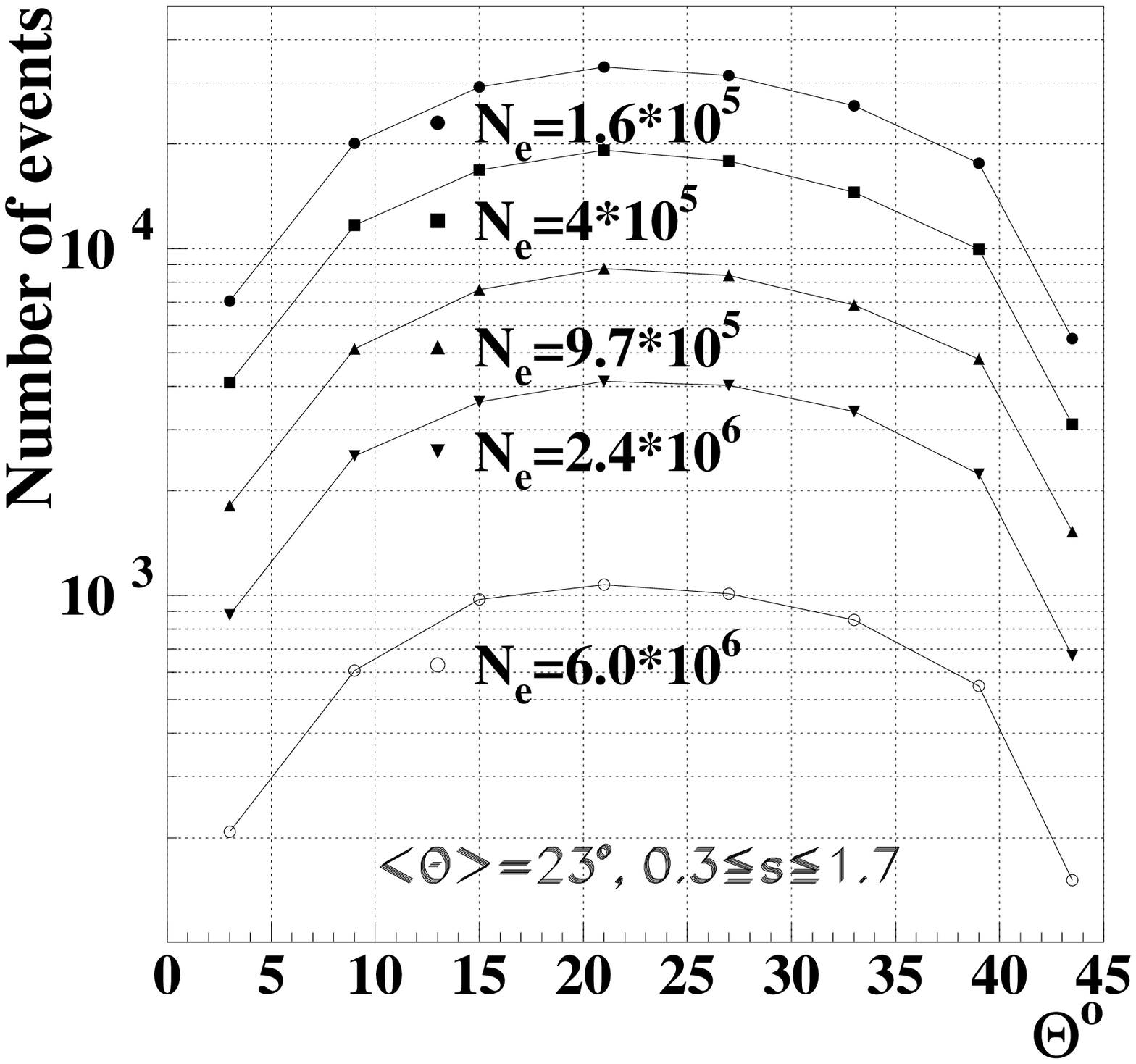,width=8.cm,height=8.cm}
\vspace{-1.5cm}
\caption{\it{Distributions of the EAS zenith angles for different 
EAS sizes.}}
\end{minipage}
\end{figure}
\begin{wrapfigure}[26]{r}{8.cm}
\vspace{-.2cm}
\epsfig{file=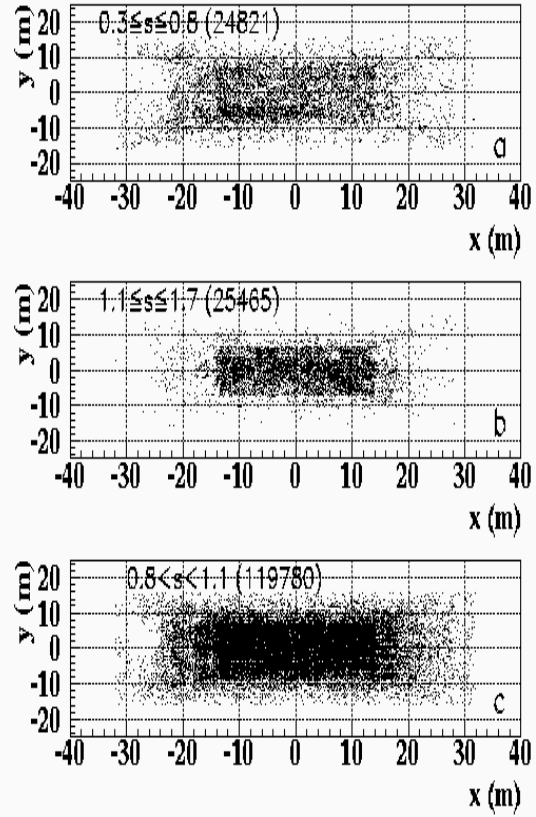,width=8.cm,height=13.cm}
\vspace{-1.7cm}
\caption{\it{Distribution of the core locations of 
different age classes of showers.}}
\vspace{-0.6cm}
\end{wrapfigure}
from the zenith angle of  
$\overline{\theta}\simeq 23^o$. For more detailed analyses the 
EAS sample is divided in three classes: "young" showers with 
$0.3 < s < 0.8$, "mature" showers with $0.8 < s < 1.1$, 
and "old" showers with $1.1 < s < 1.7\,$.\\
Figure 3 display the uniform efficiency of the age selection
of the procedures.  

\section{Age parameter distributions}
The distributions of the age parameter values for various EAS 
sizes are shown in Figure 4, displayed for different ranges of 
the zenith angles of EAS incidence.
The distributions get narrower and show decreasing variances with 
increasing $N_e$ in agreement with Ref. \cite{AH-miayke}.
This can be understood that small size showers penetrating in the 
deeper atmosphere show larger fluctuations in s.
The average age is slightly, but systematically shifted to higher 
values with increasing atmospheric depth. \\
For a consideration of the dependence of the average age from $N_e$  
and zenith angle a finer binning of the total angular range has 
been applied. As examples in Figure 5 the dependence of the mean 
age is shown for selected angular bins (representing ''vertical'', 
''inclined'' and all showers). The results are compared with 
the Norikura data \cite{AH-miayke}, which show similar tendencies, 
but shifting the global features%
\begin{figure}[h]
\begin{center}
\begin{minipage}[t]{0.32\linewidth}
\vspace{-.1cm}
\epsfig{file=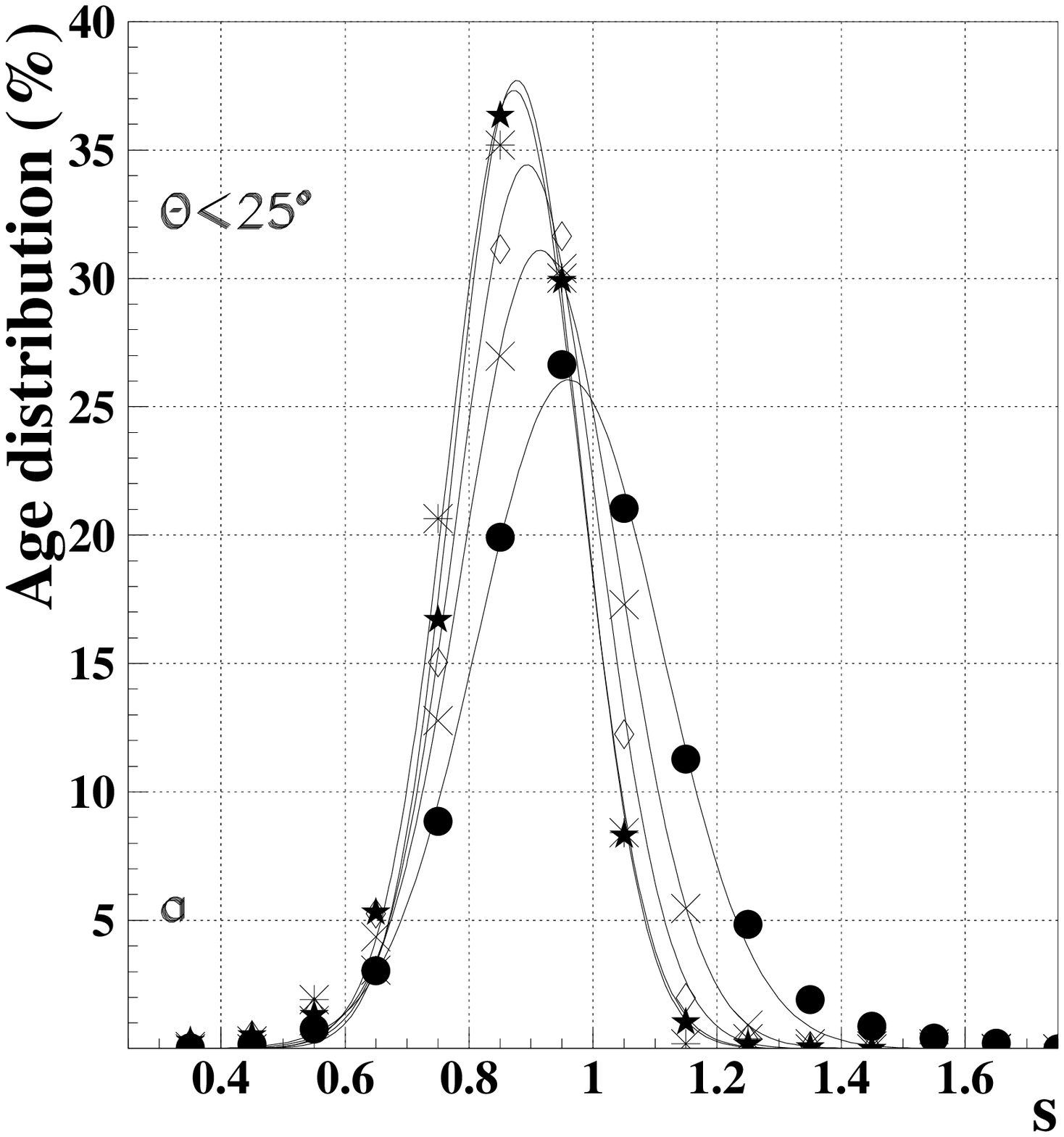,width=6.cm,height=6.cm} 
\end{minipage}
\hspace*{0.01cm}
\begin{minipage}[t]{0.32\linewidth}
\vspace{-.1cm}
\epsfig{file=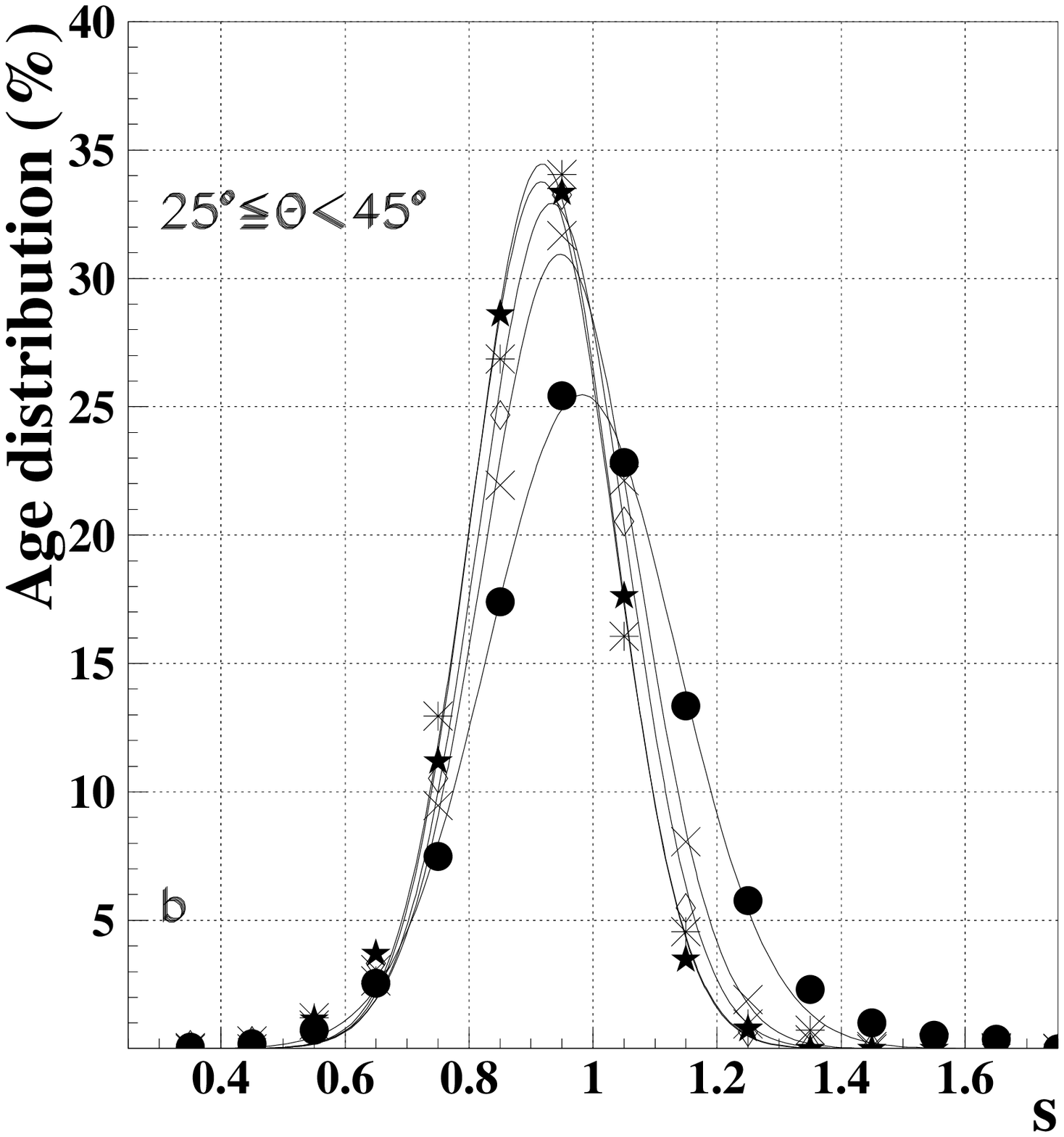,width=6.cm,height=6.cm} 
\end{minipage}
\hspace*{0.01cm}
\begin{minipage}[t]{0.32\linewidth}
\vspace{-.1cm}
\epsfig{file=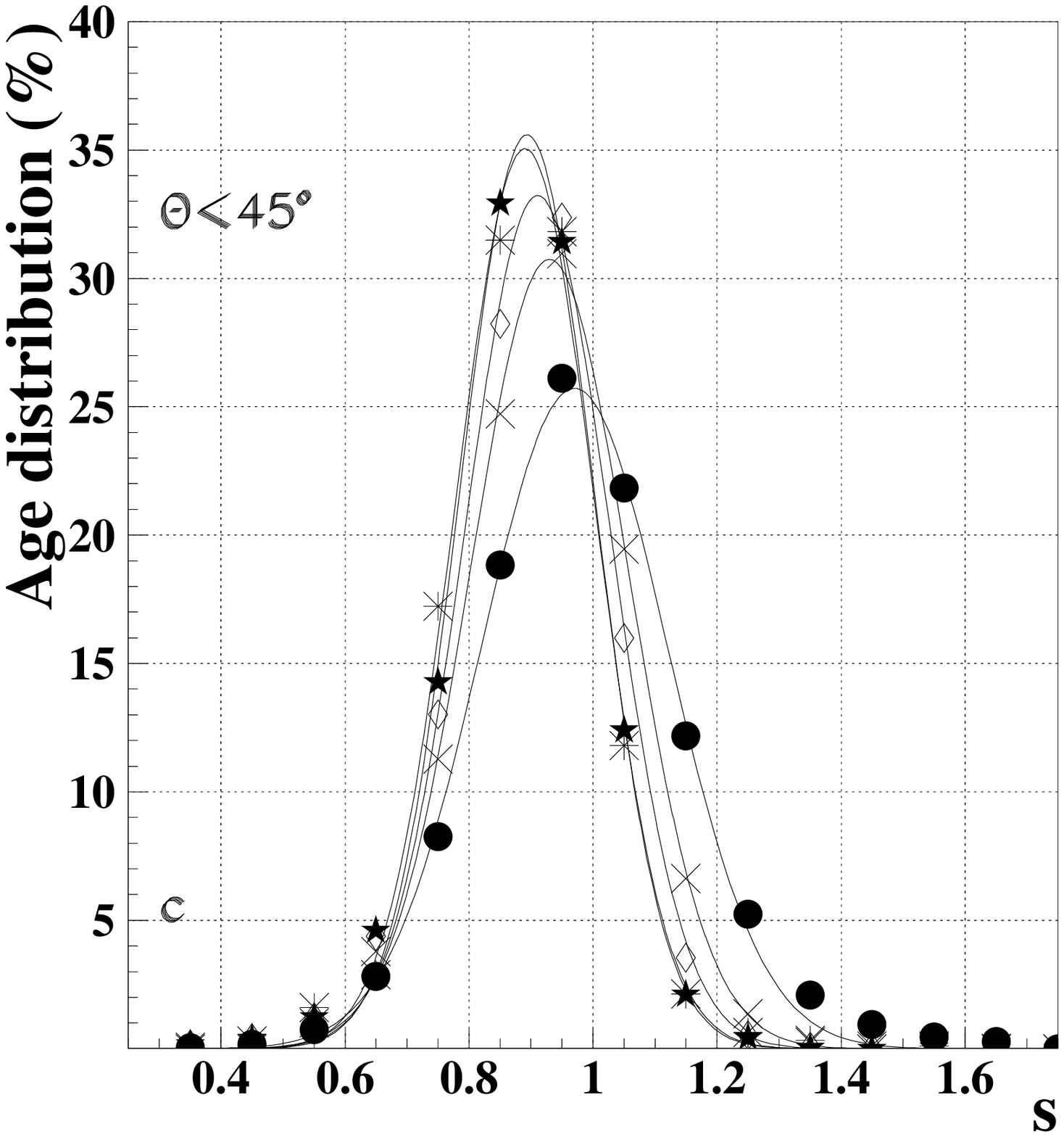,width=6.cm,height=6.cm} 
\end{minipage}
\end{center}
\vspace{-.2cm}
\caption{\it{Age parameter distributions for various EAS sizes 
and angular ranges of EAS incidence: a- vertical, 
b- inclined, c- all showers. $N_e=\bullet-1.6\cdot 10^5,
\times -4.0\cdot 10^5, \diamond-9.7\cdot10^5, \star-2.4\cdot 10^6,
\ast-6.0\cdot 10^6$.}}
\end{figure}
%

\begin{wrapfigure}[18]{l}{8.cm}
\vspace{-.7cm}
\epsfig{file=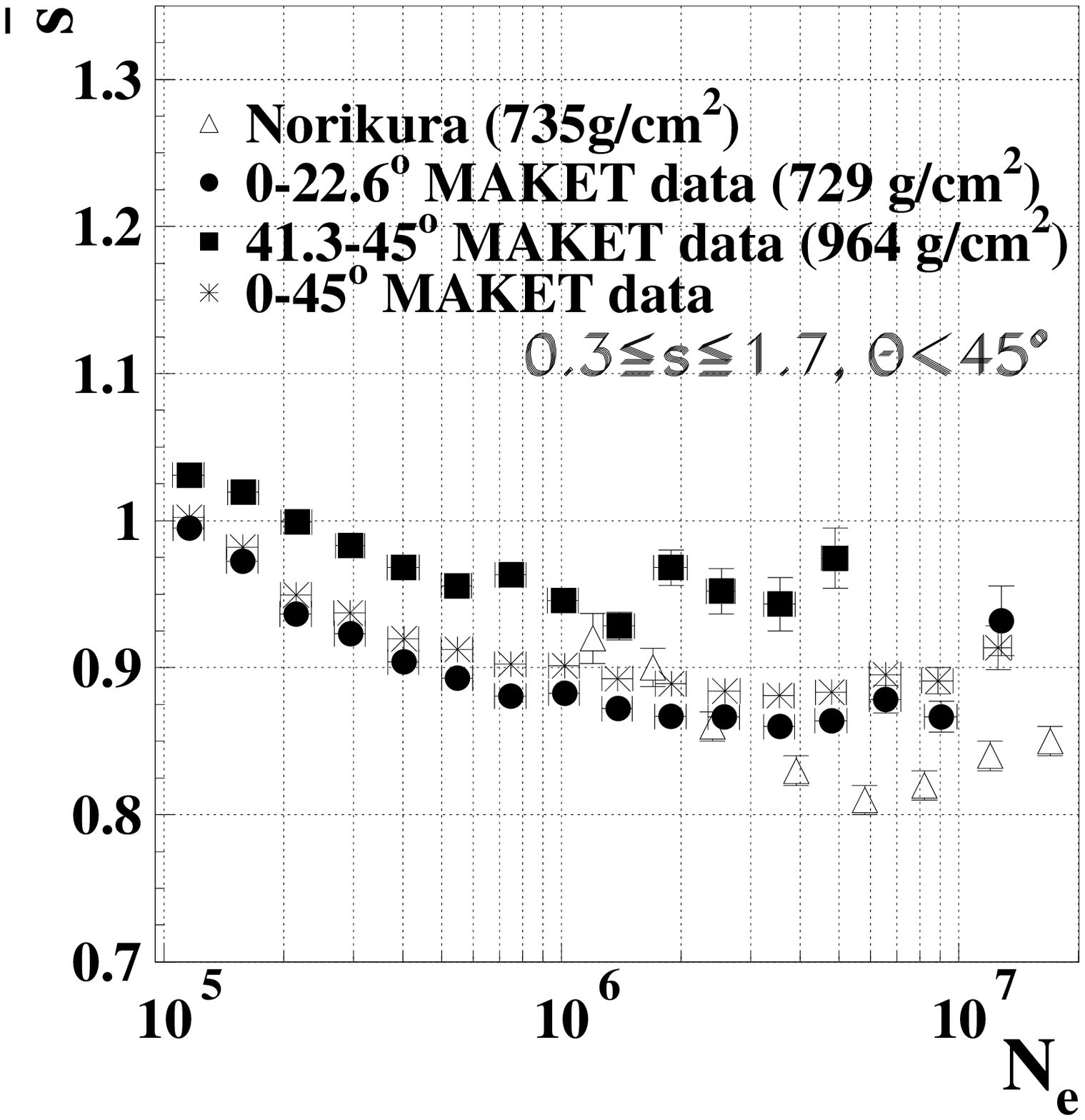,width=8.cm,height=8.cm}
\vspace{-1.cm}
\caption{\it{$\overline{s}$ dependence of the shower size.}}
\end{wrapfigure}
%

\noindent to larger and younger EAS.
It is not clear if this finding is due  methodical effects of 
different evaluation procedures in both experiments.
The results of MAKET-ANI agree with the observations 
Ref. \cite{AH-asaki}, 
if taking into account the different 
observation levels, but disagrees with the data of the MSU group 
\cite{AH-khrist}, the latter claiming an almost constant mean 
age for EAS of $N_e =10^5-10^6$. There are results of EAS 
simulations, based on the QGSJET model as generator \cite{AH-ostap}, 
which show fair agreement \cite{AH-ost}. \\
The variation of the average age is affected by the primary energy 
spectrum, by the change of the chemical composition and the 
hadronic interaction characteristics, governing the EAS development. 
As long as there is no noticeable change, the average depth of 
the shower maximum is expected to be increasing monotonously. 
Hence the shallow slope of the average age for $N_e> 10^6$ may 
indicate a faster EAS development due to an increasing 
multiplicity of the secondary production and a heavier composition, 
respectively. 

\section{EAS size spectra of different ages}
Figure 6 shows the integral size spectra for ''young'', ''mature''
and ''old'' showers for two different angular ranges of 
shower incidence. While the young and mature shower spectra 
exhibit the knee feature, a knee is not evident for old showers, 
which show obviously a different variation with the shower size. 
This behavior results also from an analysis of KASCADE  
data classified along various types of primaries by methods of 
advanced statistical analysis \cite{AH-aro}. 
The old showers are tentatively associated to iron-like showers 
with a different knee position. \\ 
The lower part of Figure 6, taken from Ref. \cite{AH-kempa}, 
where the showers have been classified by an analysis of the 
appearance of the shower core, shows a good consistency.
There are, however some differences with the Tien-Shan data 
(given in Ref.\cite{AH-nikol}). While the slopes are identical 
for mature showers and equal for old showers, the young showers 
do not display a knee in the data of Ref. \cite{AH-nikol}. 
Whether these  differences can be explained by the particular  
analysis procedures, is not yet clarified. \\
Figure 7 presents the spectra for different values of the 
age parameters and characterized by the spectral indices given 
Table 2 (extracted by the procedures of Ref. \cite{AH-sokho}). 
With increasing age values the spectral slope gets flatter 
before the knee as also evidenced by the KASCADE data 
\cite{AH-glasster}.
Old showers exhibit a quite different slope. 
\begin{figure}[h]
\begin{minipage}[t]{0.48\linewidth}
\vspace{-0.2cm}
\epsfig{file=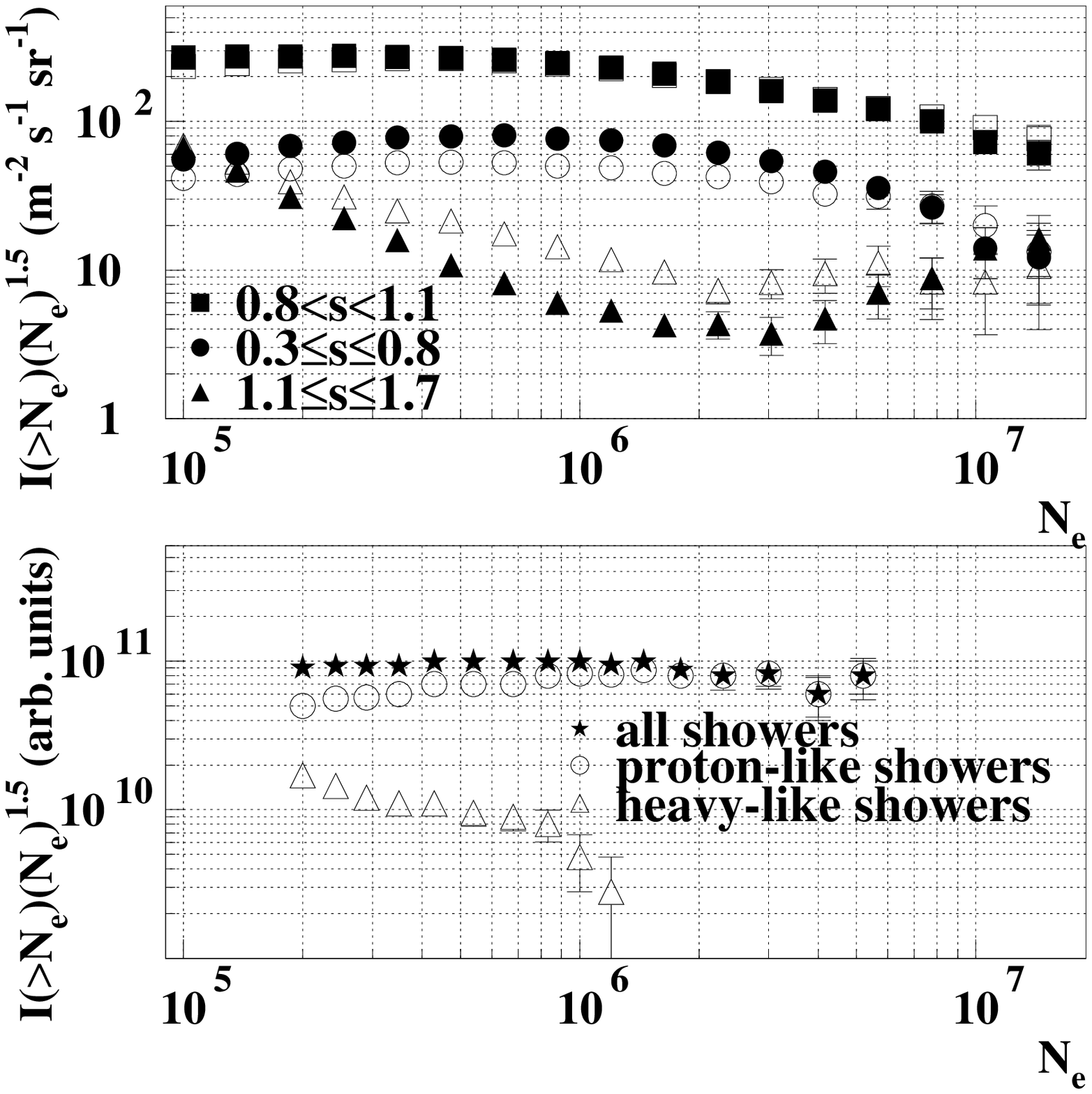,width=8.cm,height=8.cm}
\vspace{-1.cm}
\caption{\it Integral EAS size spectra for two different ranges 
of the zenith angles (closed symbols: $\theta=0^\circ -25^\circ$, 
open symbols: $\theta=25^\circ -45^\circ$ for young, mature and 
old showers. The lower part of the figure is taken from Ref. 
\cite{AH-kempa} for comparison.}
\end{minipage}
\hspace*{0.4cm}
\begin{minipage}[t]{0.48\linewidth}
\vspace{-.2cm}
\epsfig{file=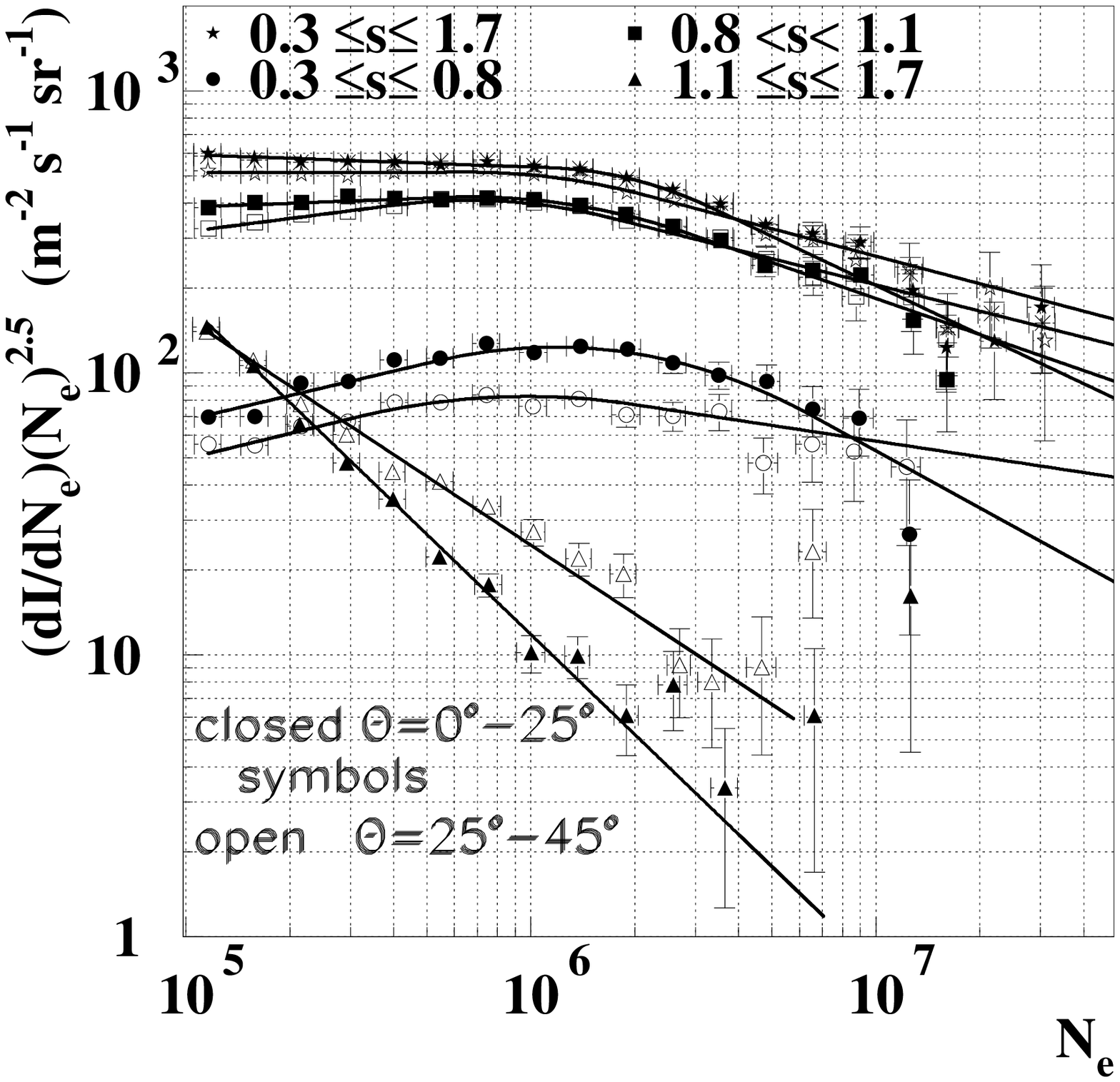,width=8.cm,height=8.cm}
\vspace{-1.cm}
\caption{\it{Differential EAS size spectra for different angular  
and age ranges.}}
\end{minipage}
\end{figure}
\begin{table}[ht]
\vspace{-0.05cm}
\caption{\it Average age values and variances for different 
zenith angles ($\theta<25^o$, $25^o\leq\theta<45^o$, $\theta<45^o$)
and EAS sizes together with the values of the parameters A  
and $s(\theta = 0)$ of the parameterization of the $\sec\theta$ 
dependence.}
\vspace{0.5cm}
\begin{center}
\begin{tabular}{|c|c|c|c|c|c|c|c|c|}
\hline
  & \multicolumn{2}{|c|}{$\theta<25^o$}
  & \multicolumn{2}{|c|}{$25^o\leq\theta<45^o$}
  & \multicolumn{2}{|c|}{$\theta<45^o$}&
  &
  \\
  \raisebox{1.5ex}[1.5ex]{$N_e$} &
  $\overline{s}$&$\sigma_s$&$\overline{s}$&
      $\sigma_s$&$\overline{s}$&$\sigma_s$&
      \raisebox{1.5ex}[1.5ex]{$A$} &  
      \raisebox{1.5ex}[1.5ex]{$s(0)$}\\
\hline

$1.6\times10^5$&$0.96$&$0.15$&$0.98$&$0.15$&$0.97$&$0.15$&$0.126\pm.002$&$0.968\pm.001$\\

$4.0\times10^5$&$0.92$&$0.13$&$0.95$&$0.13$&$0.93$&$0.13$&$0.194\pm.004$&$0.902\pm.005$\\

$9.7\times10^5$&$0.89$&$0.11$&$0.93$&$0.12$&$0.91$&$0.12$&$0.241\pm.006$&$0.872\pm.007$\\

$2.4\times10^6$&$0.88$&$0.10$&$0.92$&$0.11$&$0.89$&$0.11$&$0.274\pm.008$&$0.855\pm.009$\\

$6.0\times10^6$&$0.87$&$0.11$&$0.92$&$0.12$&$0.89$&$0.11$&$0.316\pm.2$&$0.852\pm.032$\\

$\geq10^5$&$0.93$&$0.14$&$0.96$&$0.14$&$0.94$&$0.14$&$0.161\pm.002$&$0.934\pm.001$\\
\hline
\end{tabular}
\end{center}
\end{table}
\begin{table}[ht]
\vspace{-0.05cm}
\caption{\it Spectral slopes ($dI/DN_e \propto N_e^{-\gamma}$)
and knee positions for different 
ranges of the age parameter values.}
\vspace{0.2cm}
\begin{center}
\begin{tabular}{|c|c|c|c|c|}
\hline
 $\theta$ & $s$ & $\gamma_1$ & $\gamma_2$ & $log(N_e^{knee})$ \\ 
\hline
 $0^o-25^o$& $0.3-1.7$ & $2.54\pm0.03$ & $3.08\pm0.03$ & $6.30$\\
 $ $& $0.8-1.1$ & $2.45\pm0.03$ & $2.92\pm0.07$ & $6.13$\\
 $ $ & $0.3-0.8$ & $2.21\pm0.03$ & $3.17\pm0.14$ & $6.31$\\
 $ $ & $1.1-1.7$ & $3.68\pm0.08$ & &  \\
\hline
 $25^o-45^o$ & $0.3-1.7$ & $2.50\pm0.02$ & $2.82\pm0.04$ & $6.08$\\
 $ $ & $0.8-1.1$ & $2.34\pm0.03$ & $2.81\pm0.05$ & $5.93$\\
 $ $ & $0.3-0.8$ & $2.20\pm0.02$ & $2.70\pm0.07$ & $5.91$\\
 $ $ & $1.1-1.7$ & $3.31\pm.07$ & &  \\
\hline
\end{tabular}
\end{center}
\end{table}

\section{Variation of the age with the observation depth}
Figure 8 shows the dependence of the mean age $\overline{s}(\theta)$ of 
particular EAS sizes from the zenith angle $\theta$, as 
linear dependence from  $\sec\theta$. \\
The parameters $\overline{s}(0)$ and A, adjusted to the $\sec\theta$ 
dependence are given in Table 1. \\
With increasing $N_e$ the slope A increases while $\overline{s}(0)$ is 
decreasing. There is a good agreement with the%
\begin{wrapfigure}[20]{r}{8.cm}
\vspace{-.7cm}
\epsfig{file=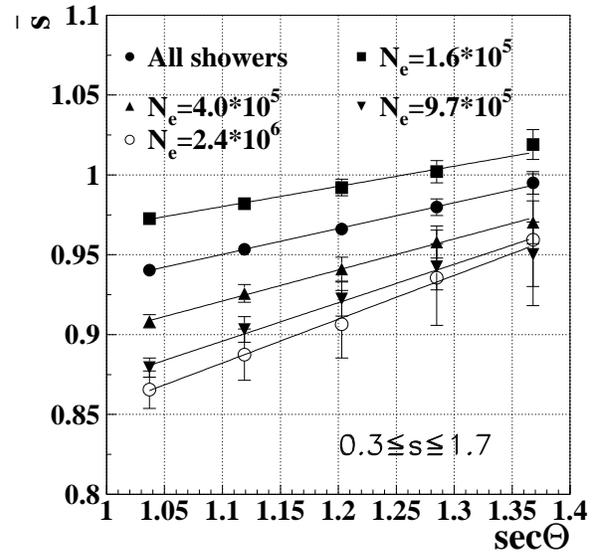,width=8.5cm,height=8.5cm}
\vspace{-1.cm}
\caption{\it{The dependence of the mean age $\overline{s}$ from the 
zenith angle of EAS incidence for various shower sizes.}}
\end{wrapfigure}
values of Ref. \cite{AH-miayke} obtained for $N_e =2.4 \cdot 10^6$. 
The values averaged over all EAS sizes are $A=0.161\pm0.002$
and $\overline{s}(0) =0.934 \pm 0.001$. With the approximate relation 
$\sec\theta=X/X_v$ where $X_v$ is depth of the observation 
level and X the transverse atmospheric thickness (grammage) 
A can be related to the change $d\overline{s}/dX$ of the average age 
with X. With the average value of A inferred from the data for 
the observation level $X_v = 700\,$g/cm$^2$ a value $d\overline{s}/dX = 
2.3 \cdot 10^{-4}\,$cm$^2$/g. This result can be compared 
with $d\overline{s}/dX = 3.4 \cdot 10^{-4}\,$cm$^2$/g given in 
\cite{AH-miayke}. A compilation \cite{AH-miayke} of the data 
from the literature yields a range $d\overline{s}/dX = (1.9-4.3) \cdot 
10^{-4}\,$cm$^2$/g. Associating the depth of the shower 
maximum $X_m$ with  $\overline{s}=1$,
we reach the relation 
\begin{equation}
\overline{s} -1 ={\frac{d\overline{s}}{dX}}\cdot (X-X_m),
\end{equation}
Thus an evaluation of the $N_e$ dependence of $\Delta X=(X-X_m)$
carries some information about the elongation rate, as already 
indicated by Linsley \cite{AH-linsley}.

\section{Concluding Remarks}
The present results deduced from the data of the MAKET-ANI 
array are in good agreement with theoretical expectations. The analyses reveal that: \\
$\bullet$ Average age parameter gradually decreases with increasing shower size
from $10^5$ to $10^6$, and for $N_e>10^6$ it becomes almost constant.\\
$\bullet$ The knee of "young" showers is sharper than knee of the all particle spectra.\\ 
$\bullet$  The size spectra classified by different ages show 
different attenuation. \\
$\bullet$  The change of age parameter with the zenith angle of 
EAS incidence can be related to the change of the EAS maximum 
with $N_e$.

\vspace*{3.5ex}
{\noindent \bf \Large Acknowledgment \\}
\vspace*{-.5ex}

	{\it The report is based on scientific results of the 
	ANI collaboration. The MAKET-ANI installation on Mt.Aragats 
	has been setup as a collaboration project of the 
Yerevan Physics Institute (Armenia)  and the Lebedev Physics Institute (Moscow).
The continuous contributions and assistance of the Russian colleagues in 
operating of the  installation and in the data analyses are gratefully 
acknowledged. In particular, we thank Prof. S. Nikolski and Dr. V. Romakhin 
for their encouraging interest to this work and useful discussions. \\
	We would like to thank Prof. Dr. H. Rebel pointing on the 
importance of the elongation rate estimation, Dr. A. Haungs and Dr. Kh. Sanosyan 
for useful remarks.
Corrections and suggestions made by Dr. S. Ostapchenko are highly appreciated.
The assistance of the Maintenance Staff of the Aragats Cosmic Ray
Observatory in operating the MAKET-ANI installation is highly appreciated.
The work has been partly supported by the ISTC project A116.}

\setcounter{section}{0}
\setcounter{footnote}{0}
\setcounter{figure}{0}
\setcounter{table}{0}
\newpage
\clearpage


\begin{thebibliography}{99}
\renewcommand{\baselinestretch}{0.1}
\parskip0.ex
\bibitem{AH-miayke}
S. Miayke et al., Proc. 16$^{th}$ ICRC (Kyoto) {\bf 13} (1979) 171\\
S. Miayke et al., Proc. 17$^{th}$ ICRC (Paris) {\bf 11} (1981) 293\\
B.S. Acharya et al., Proc. 17$^{th}$ ICRC (Paris) {\bf 9} (1981) 162
\bibitem{AH-kempa}
J. Kempa and M. Samorski, J.Phys. G: Nucl. Part. Phys. {\bf 24} (1998) 1039
\bibitem{AH-khrist}
G.B. Khristiansen et al., Proc. AS USSR, Phys. {\bf 35} (1971), 2107 (in Russian)\\
G.B. Khristiansen et al., Proc. 17$^{th}$ ICRC (Paris) {\bf 6} (1981) 39\\ 
N.N. Kalmykov et al., Proc. 25$^{th}$ ICRC (Durban) {\bf 6} (1997) 277 
\bibitem{AH-avak}
V.V. Avakian et al., Soviet J. Nucl. Phys. {\bf 56} (1993) 183 (in Russian) 
\bibitem{AH-gagik}
G.V. Gharagyozyan for the ANI collab., Proc. of the Workshop ANI 98, eds.
A.A. Chilingarian, H.Rebel, M. Roth, M.Z. Zazyan, FZKA 6215,
Forschungszentrum Karlsruhe 1998, p.51
\bibitem{AH-laura}
L.G. Melkumyan for the ANI collab., ANI Workshop  1999, these proceedings
\bibitem{AH-asaki}
K. Asakimori et al., Proc. 17$^{th}$ ICRC (Paris) {\bf 11} (1981) 301\\
T. Hara et al., Proc. 17$^{th}$ ICRC (Paris) {\bf 6} (1981) 52\\
K. Asakimori, Proc. 21$^{st}$ ICRC (Adelaide) {\bf 3} (1990) 129
\bibitem{AH-ostap}
N.N. Kalmykov, S.S. Ostapchenko and A.J. Pavlov, Nucl. Phys. B {\bf 52R}
(1997) 17
\bibitem{AH-ost}
S.S. Ostapchenko, private communication
\bibitem{AH-aro}
A. Vardanyan et al. - KASCADE collaboration, these proceedings
\bibitem{AH-nikol}
S.I. Nikolsky, Proc. 25$^{th}$ ICRC (Durban) {\bf 6} (1997) 105
\bibitem{AH-sokho}
S.H. Sokhoyan et al., these proceedings
\bibitem{AH-glasster}
R. Glasstetter et al. - KASCADE collaboration, Proc. 25$^{th}$ ICRC (Durban) {\bf 6} 
(1997) 157
\bibitem{AH-linsley}
J. Linsley, Proc. 15$^{th}$ ICRC (Plovdiv) {\bf 12} (1977) 89 
\end{thebibliography}
\end {document}